\documentclass[%
 reprint,
 amsmath,amssymb,
 aps,prl
]{revtex4-2}
\makeatletter

\usepackage{graphicx}
\usepackage{dcolumn}

\usepackage{bm}
\def\@make@capt@title#1#2{%
  \vspace{10pt}%
  \begingroup
    \small
    \setlength{\parindent}{0pt}%
    \setlength{\leftskip}{0pt}%
    \setlength{\rightskip}{0pt}%
    \noindent
    #1\@caption@fignum@sep#2\par
  \endgroup
}
\makeatother

\usepackage{soul}
\usepackage{xcolor}

\begin{document}

\preprint{APS}

\title{Explicit Solvents Reverse Temperature Dependence of Phase Behavior in Charged Fluids
}

\author{Pierre J. Walker}
 \email{pjwalker@caltech.edu}
\author{Zhen-Gang Wang}%
 \email{zgw@caltech.edu}
\affiliation{Division of Chemistry and Chemical Engineering, California Institute of Technology, Pasadena, CA 91125, United States
}%

\begin{abstract}
In most theories of electrolyte and polyelectrolyte solutions, the solvent is treated implicitly as a dielectric continuum. While this approximation has enabled major advances, it neglects coupling between ionic and dipolar degrees of freedom. Using molecular dynamics simulations and the mean-spherical approximation, we show that this coupling qualitatively alters electrostatically driven phase behaviour. At finite ion concentrations, many-body ion–dipole correlations reverse the temperature dependence of effective electrostatic interactions, leading to upper-critical solution temperature (UCST) behaviour in direct contrast to the lower-critical solution temperature (LCST) behaviour predicted by implicit-solvent electrostatic theories. This phase reversal arises from ion-induced suppression of local dipolar screening, a collective effect inaccessible to pairwise implicit-solvent descriptions.
\end{abstract}

\maketitle


Much of our understanding of electrolyte and polyelectrolyte solutions is based on the use of implicit-solvent models, in which the solvent is represented as a dielectric continuum, with polarization and dipolar degrees of freedom subsumed into a single parameter, the dielectric constant, which rescales Coulomb interactions as\cite{maxwellTreatiseElectricityMagnetism1873}
\begin{equation}
\label{eq:coul}
\beta\phi_\mathrm{el}(r)=\frac{q_iq_j}{4\pi\epsilon_0\epsilon_\mathrm{r}k_\mathrm{B}Tr}
= z_i z_j\frac{l_\mathrm{B}}{r},
\end{equation}
where $q_i$ and $q_j$ are the particle charges, $z_i$ and $z_j$ their valencies, and $l_\mathrm{B}=e^2/(4\pi\epsilon_0\epsilon_\mathrm{r}k_\mathrm{B}T)$ is the Bjerrum length, a measure of the strength of electrostatic interactions relative to thermal energy. This framework played a central role in the early classical theories of Gouy--Chapman\cite{gouyConstitutionChargeElectrique1910,chapmanLIContributionTheory1913}, Debye--H\"uckel\cite{debyeZurTheorieElektrolyte1923}, Wagner\cite{wagnerTheorieAlterungNiederschlaegen1961}, and Onsager\cite{onsagerZurTheorieElectrolyte1926,onsagerZurTheorieElectrolyte1927}. These developments established many of the foundational concepts in the physics of ionic solutions, including electrostatic screening, the structure of the electrical double layer, and ionic correlations. More contemporary treatments incorporate the finite hard-core size of ions\cite{blumMeanSphericalModel1975,kusalikMolecularTheoryAqueous1988}, leading to the standard primitive model, whose analysis constitutes a central component of liquid-state theory and underpins much of the modern computer simulation of ionic and electrolyte systems. 


\begin{figure}[ht!]
    \includegraphics[width=\linewidth]{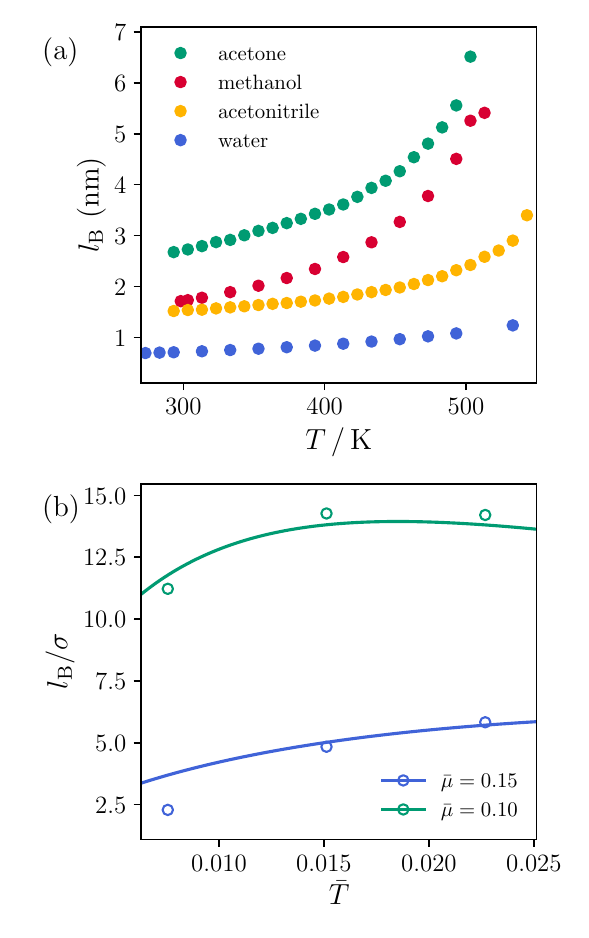}
    \vspace{-1.2cm}
    \caption{(a) Bjerrum length $l_\mathrm{B}$ as a function of temperature for water, methanol, acetonitrile, and acetone, computed from experimental static permittivity data (obtained from the DETHERM databank\cite{iltenDETHERMThermophysicalProperty1991}). (b) Bjerrum length $l_\mathrm{B}/\sigma$ as a function of dimensionless temperature $\bar{T}$ for the hard-sphere dipolar fluid at two dipole strengths $\bar{\mu}$; open circles denote MD simulation results and solid lines denote MSA-ID theory predictions.
    }
    \label{fig:1}
    \vspace{-0.5cm}
\end{figure}

\begin{figure*}[!ht]
    \includegraphics[width=\linewidth]{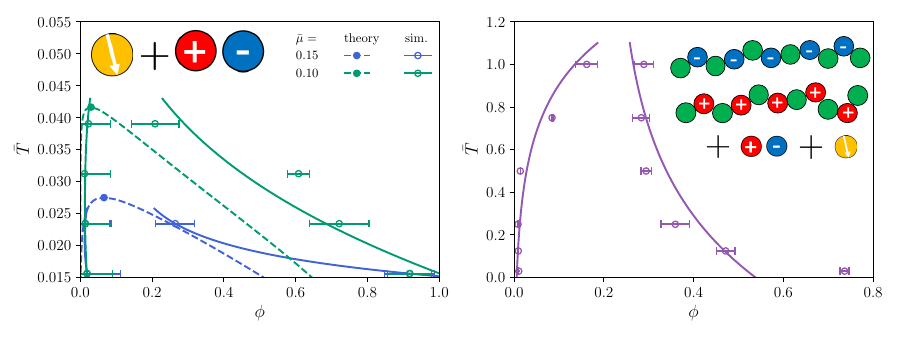}
    \vspace{-1.2cm}
    \caption{Co-existing volume fractions of the charged components, $\phi$, as a function of dimensionless temperature $\bar{T}$ from theory (dashed lines, solid circles) and MD simulations (solid lines, open circles) for (a) monomeric ions at varying dipole strengths $\bar{\mu}$ and (b) oppositely charged polyelectrolyte chains with counterions and solvent dipoles with strength $\bar{\mu}=0.15$. Filled symbols denote critical points predicted by theory. The solid lines are included as visual guides for the phase boundary.}
    \label{fig:2}
    \vspace{-0.5cm}
\end{figure*}

For a fixed dielectric constant, increasing temperature decreases the Bjerrum length and thereby weakens electrostatic interactions. In polar solvents with large molecular dipole moments, however, thermal disruption of dipolar correlations can reduce the dielectric constant sufficiently to yield a net increase in the Bjerrum length (Fig.~\ref{fig:1}a). Consequently, electrostatic interactions can become stronger upon heating, in contrast to the behavior of most other intermolecular forces. This reasoning leads naturally to the expectation that electrostatic-correlation-induced phase separation in polar liquids should be favored at elevated temperatures, resulting in lower critical solution temperature (LCST) behavior in electrolyte and polyelectrolyte systems. Indeed, the temperature dependence of $\epsilon_\mathrm{r}$ has been invoked to explain experimentally observed LCST behavior in aqueous polyelectrolyte complex coacervation\cite{aliLowerCriticalSolution2019,adhikariLowerCriticalSolution2019,ylitaloElectrostaticCorrelationsTemperatureDependent2021}. However, this picture is inconsistent with the increasing solubility of many simple salts with temperature and fails to explain the upper critical solution temperature (UCST) behavior observed in other polyelectrolyte systems\cite{kantasharkerUpperCriticalSolution2019,ohUpperCriticalSolution2024,liSaltInducedLiquidLiquid2021}.

A fundamental limitation of the implicit-solvent framework is its neglect of coupling between ionic and dipolar degrees of freedom. It is well established that the addition of salt can produce a dielectric decrement\cite{shockMolecularDynamicsSimulations2023,maribo-mogensenModelingDielectricProperties2013a,sealDielectricDecrementAqueous2019,gaoAnalyticalExpressionDielectric2020,buyukdagliExplicitSolventTheory2022}, arising primarily from orientational saturation of solvent dipoles within ionic solvation shells. More generally, direct ion--dipole interactions generate solvent-mediated many-body interactions between ions that cannot be captured by the pairwise Coulombic description of Eq.~\ref{eq:coul}. As we show below, these effects can reverse the temperature dependence predicted from the implicit-solvent picture and give rise to UCST phase behavior.

To elucidate this mechanism, we adopt a minimal model in which solvent molecules are represented as  repulsive spherical particles carrying point dipoles, while ions are repulsive spherical particles carrying a unit charge $e$. All species share a common diameter $\sigma$, and the total number density is fixed at $\rho\sigma^3=0.8$. We define a reduced temperature,
\begin{equation}
\bar{T}=\frac{4\pi\epsilon_0 k_\mathrm{B}T\sigma}{e^2},
\end{equation}
corresponding to the inverse vacuum Bjerrum length scaled by $\sigma$, and a reduced dipole strength,
\begin{equation}
\bar{\mu}=\frac{\mu}{e\sigma},
\end{equation}
where $\mu$ is the solvent dipole moment. For a bead diameter of $0.4$\,nm, $\bar{T}\approx0.008$ corresponds approximately to $300$\,K, while $\bar{\mu}=0.15$ is comparable to the dipole moment of water. As shown in Fig.~1b, over the temperature range shown this minimal dipolar-solvent model reproduces the increase in Bjerrum length with temperature characteristic of the polar solvents in Fig.~1a. For the polyelectrolyte systems, we use alternating charged and neutral beads to represent the spatial separation of charged groups along a polymer backbone of length $N=11$, along with oppositely charged counterions.  To approximately account for electronic polarizability, all point charges are reduced by a factor of $1/\sqrt{2}$, while the solvent dipole moments are left unchanged. This is equivalent, for interactions between point charges, to introducing a temperature-independent background permittivity of $\epsilon_\infty=2$, as commonly employed in simulations of charged molecular systems\cite{duboue-dijonPracticalGuideBiologically2020}.

To determine the phase behavior while explicitly accounting for ion--dipole coupling, we employ two complementary approaches: molecular dynamics simulations and, for the monomeric ion systems, the mean-spherical approximation for ion--dipole mixtures (MSA-ID)\cite{blumAnalyticalSolutionMean1987,simoninSolutionMeansphericalApproximation2020,simoninFullSolutionMean2021}. All theoretical calculations are carried out using the Clapeyron.jl package\cite{walkerClapeyronjlExtensibleOpenSource2022}. Simulations are performed using LAMMPS in a slab geometry\cite{thompsonLAMMPSFlexibleSimulation2022,alonsoExplicitSolventModel2026}.

The resulting coexistence volume fractions are shown in Fig.~\ref{fig:2} for monomeric ions (Fig.~\ref{fig:2}a) and salt-free polyelectrolyte chains (Fig.~\ref{fig:2}b). Strikingly, all systems exhibit UCST behavior. For the $\bar{\mu}=0.15$ monomeric-ion system, this occurs despite the increase in the dilute-solvent Bjerrum length over the overlapping temperature range shown in Fig.~1b. The full phase behavior therefore cannot be inferred from the temperature dependence of the dilute-solvent Bjerrum length alone. Examination of the Bjerrum length in the coexisting phases (see ESI) reveals markedly different thermal responses. In the charge-dilute phase of Fig.~2a, where the dielectric response is dominated by the dipolar solvent, the Bjerrum length increases with temperature. In the charge-dense phase, solvent depletion suppresses the strongly temperature-dependent orientational contribution to the dielectric response, such that the Bjerrum length instead decreases with temperature. Electrostatic interactions therefore weaken upon heating in the charge-dense phase, consistent with the observed UCST phase behavior.

The same physical picture explains the trends across the phase diagrams. Decreasing the solvent dipole strength reduces dielectric screening and strengthens electrostatic interactions, shifting the critical temperature upward (Fig.~\ref{fig:2}a).  Replacing monomeric ions with polyelectrolyte chains shifts the coexistence boundary toward higher temperatures and lower critical volume fractions (Fig.~2b). This shift reflects both the stronger collective electrostatic interactions generated by multiple charges along each chain and the substantial reduction in translational entropy associated with polymerization. In both cases, the phase diagrams remain strongly asymmetric, skewed toward the charge-dilute region, reflecting the long-range and many-body character of electrostatic interactions\cite{fisherCriticalityIonicFluids1993,luijtenUniversalityClassCriticality2002,orkoulasMonteCarloStudy2003,singElectrostaticControlBlock2014,walkerAnomalousPhaseBehavior2025}.

The phase-dependent Bjerrum lengths are therefore consistent with the observed UCST behavior, but this correspondence does not establish its microscopic origin. The dielectric constant is a bulk response function and does not uniquely determine the effective interaction experienced by individual charged species. We therefore calculate potentials of mean force (PMFs) between oppositely charged species\cite{chenDrivingForcePathway2022,shiDeconstructingFreeEnergies2017,briniHowWatersProperties2017,oliverWhenB2Not2025} to determine directly how the local interactions evolve with temperature and with the surrounding ionic environment.

\begin{figure}[ht!]
    \includegraphics[width=\linewidth]{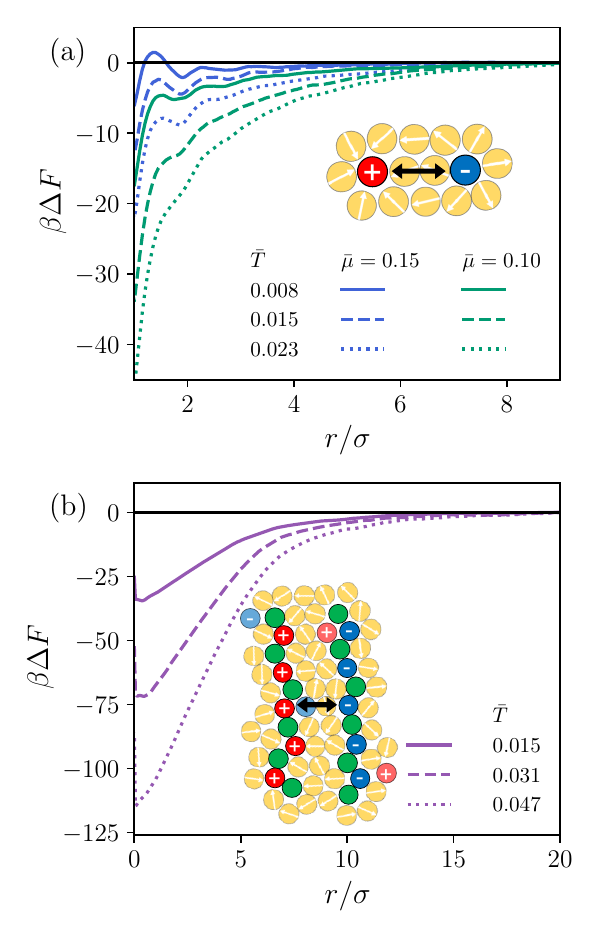}
    \vspace{-1.2cm}
    \caption{(a) PMF ($\beta\Delta F$) between two oppositely charged monovalent ions as a function of separation $r/\sigma$ for varying $\bar{\mu}$ and $\bar{T}$. (b) PMF between two oppositely charged polyelectrolyte chains of degree of polymerization $N=11$ as a function of center-of-mass separation $r/\sigma$ at varying $\bar{T}$ (with $\bar{\mu}=0.15$).}
    \label{fig:3}
    \vspace{-0.5cm}
\end{figure}

We compute the PMF between oppositely charged monovalent ions and between oppositely charged polyelectrolyte chains at varying temperatures and dipole strengths using the adaptive biasing force (ABF) method\cite{darveAdaptiveBiasingForce2008,heninExploringMultidimensionalFree2010} implemented in the ColVars module\cite{fiorinUsingCollectiveVariables2013} within an isotropic box. The resulting PMFs are shown in Fig.~\ref{fig:3}. Across the range of dipole strengths examined, pair attractions become stronger with increasing temperature. Consistent with Eq.~\ref{eq:coul} and prior work\cite{varnerEntropicOriginIonic2023}, decreasing the dipole strength lowers the dielectric response and deepens the attractive well. At $\bar{\mu}=0.15$, the monomeric-ion PMF additionally exhibits short-ranged oscillations associated with solvent-shell structure once dipolar screening substantially weakens the direct ion--ion attraction. The same qualitative thermal response is observed between oppositely charged polyelectrolyte chains (Fig.~\ref{fig:3}b), for which the attraction likewise strengthens with increasing temperature.


Taken in isolation, these PMFs appear to support the implicit-solvent picture: increasing temperature reduces the dielectric screening sufficiently to strengthen the attraction between charged species. Yet this trend is opposite to the UCST phase behavior observed in Fig.~\ref{fig:2}, where electrostatic stabilization of the charge-dense phase must weaken upon heating. The isolated pair interaction therefore cannot, by itself, explain the bulk phase behavior. The missing ingredient is the concentration dependence of the electrostatic environment, which introduces collective ion--ion and ion--dipole correlations even at relatively low charge concentrations.

\begin{figure}[ht!]
    \includegraphics[width=\linewidth]{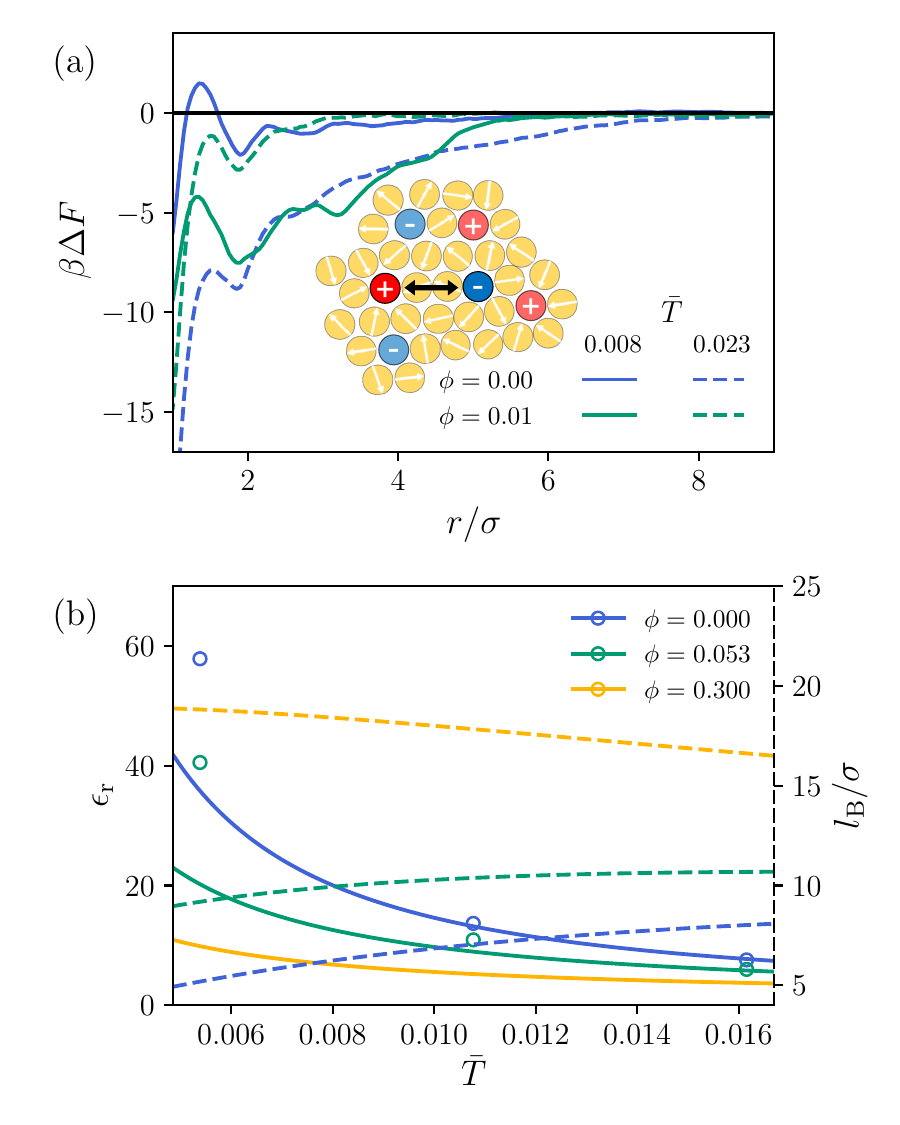}
    \vspace{-1.2cm}
    \caption{(a) PMF $\beta\Delta F$ between two oppositely charged monomeric ions ($\bar{\mu}=0.15$) at ion volume fractions $\phi=0.00$ and $\phi=0.01$ for two temperatures $\bar{T}$, illustrating the reversal in temperature dependence upon addition of ions. (b) Bjerrum length $l_\mathrm{B}/\sigma$ as a function of $\bar{T}$ for three ion volume fractions $\phi$; open circles denote MD simulation results and solid lines denote MSA-ID theory predictions.}
    \label{fig:4}
    \vspace{-0.5cm}
\end{figure}

To test this concentration dependence directly, we repeat the ion--ion PMF calculations at finite ion concentration. As shown in Fig.~\ref{fig:4}a, introducing a small ion fraction ($\phi=0.01$, corresponding to approximately $0.2$,M for $\sigma=0.4$,nm) produces two distinct effects. First, ionic screening shortens the interaction range relative to the infinite-dilution limit\cite{debyeZurTheorieElektrolyte1923,singIonCorrelationInducedPhase2013,walkerAnomalousPhaseBehavior2025}. Second, and more importantly, the thermal response reverses: the attractive interaction now weakens with increasing temperature. Thus, the same pair of ions exhibits qualitatively different temperature dependence depending on the surrounding ion concentration. This reversal is consistent with the concentration-dependent dielectric response shown in Fig.~\ref{fig:4}b. Addition of ions reduces the dielectric constant through disruption of dipolar correlations by strong local ionic fields\cite{shockMolecularDynamicsSimulations2023,maribo-mogensenModelingDielectricProperties2013a,sealDielectricDecrementAqueous2019,gaoAnalyticalExpressionDielectric2020,buyukdagliExplicitSolventTheory2022}. It also weakens the temperature dependence of $\epsilon_\mathrm{r}$, as the orientational response of the dipolar solvent is increasingly suppressed.   While part of this trend is reflected in the bulk Bjerrum length inferred from the macroscopic dielectric response (Fig.~4b), the PMF reverses its temperature dependence at ion concentrations far below those at which the bulk Bjerrum length does so. The bulk dielectric response therefore captures part of the qualitative trend but not the concentration scale at which the microscopic interactions change. This discrepancy reflects the distinction between the macroscopic dielectric response and the local solvent environment surrounding the ions: dipolar correlations within ionic solvation shells are perturbed much more strongly than the bulk solvent\cite{boothDielectricConstantWater1951,blosseyFieldTheoryStructured2022,blosseyContinuumTheoriesStructured2022,sahinNonlocalNonlinearElectrostatics2014,buyukdagliMicroscopicFormulationNonlocal2013,buyukdagliExplicitSolventEffects2022,buyukdagliExplicitSolventTheory2022}. The resulting ion--dipole coupling generates spatially-varying, concentration-dependent many-body interactions that cannot be represented by a pairwise Coulomb potential with a uniform dielectric constant.

In essence, the mechanism is a crossover from solvent-controlled electrostatics at infinite dilution to ion-induced suppression of local dipolar screening at finite concentration, reversing the thermal response of electrostatic interactions and ultimately producing UCST phase behavior.

The elevated critical temperatures in Fig.~\ref{fig:2} and the depth of the PMFs in Figs.~\ref{fig:3} and \ref{fig:4} reflect deliberate simplifications of the model. One such simplification is the assumption of a constant reduced density, $\rho\sigma^3=0.8$.  Real liquids undergo thermal expansion, coupling changes in density to changes in the dielectric response. The resulting decrease in density weakens cohesive interactions and can substantially lower the critical temperature, particularly for long-ranged electrostatic interactions. Thermal expansion also modifies the dielectric response itself: for water under isobaric conditions, the temperature- and density-dependent dielectric constant leads to a much stronger variation in the Bjerrum length than in our constant-density model, with $l_\mathrm{B}$ reaching approximately $20\sigma$ and becoming nonmonotonic over a comparable temperature range\cite{uematsuStaticDielectricConstant1980}. Thus, neither the temperature dependence of $l_\mathrm{B}$ nor the absolute critical temperatures in the present model should be interpreted quantitatively. Fixing $\rho\sigma^3=0.8$ deliberately removes this density-driven weakening, allowing the role of ion--dipole coupling to be isolated.

A more substantial simplification is the representation of polyelectrolytes as assemblies of strongly interacting charged beads. Rather than providing a quantitative description of a specific material, this minimal representation isolates electrostatics as the dominant driving force and thereby exposes its intrinsic thermodynamic tendency. Because ion--ion interactions are strong, assembling many charged groups along a polymer backbone naturally shifts phase separation to high temperatures. In real systems, both the solvent and polymer contain electronically polarizable groups and, in many cases, permanent dipoles, such that the dielectric response contains both weakly temperature-dependent electronic and strongly temperature-dependent orientational contributions. In our polyelectrolyte simulations, the former is approximately incorporated by rescaling all charges by $1/\sqrt{2}$, corresponding to a background electronic permittivity of $\epsilon_\infty=2$\cite{duboue-dijonPracticalGuideBiologically2020}. This rescaling preserves both the concentration-dependent reversal of the thermal response in Fig.~\ref{fig:4} and the resulting UCST phase behavior.

 Interestingly, the balance between electronic and orientational contributions to the dielectric response also provides a useful criterion for the applicability of an implicit-solvent description. When dielectric screening is dominated by electronic polarizability, its temperature dependence is weak and can be approximately represented by a constant background permittivity, or equivalently by a rescaling of the charges. Under such conditions, an implicit-solvent treatment can remain appropriate\cite{kimOrganicSolventDispersion2008,mengEffectMixedSolvents2020,danielsenComplexationConjugatedPolyelectrolyte2019,leeComplexCoacervationPolymerized2021,wangPhaseSeparationPositively2025}. Importantly, however, a large orientational contribution does not by itself invalidate the continuum description. For example, in dilute aqueous electrolyte solutions, the strongly perturbed solvent within the first solvation shell may be incorporated into an effective hydrated ion, leaving the surrounding solvent sufficiently weakly perturbed that its response can be described by a bulk dielectric constant, as in Debye--H\"uckel theory. The implicit-solvent approximation becomes problematic when ion--solvent coupling produces substantial nonlinear or concentration-dependent reorganization beyond this local solvation environment, such that the dielectric response experienced by interacting ions differs qualitatively from the bulk linear response. It is under these conditions that reducing the solvent to a uniform dielectric background can fail.  

More broadly, these results illustrate both the utility and the limitations of reducing a complex charged system to a single dominant interaction. Within the minimal model considered here, electrostatic interactions alone favor UCST behavior over a broad temperature range. The coexistence of LCST and UCST behavior across experimentally studied polyelectrolyte systems can then be understood as arising from competition between this electrostatic tendency and additional interactions, including hydrophobic interactions and hydrogen bonding, whose temperature dependences may favor or oppose phase separation over experimentally accessible temperature ranges\cite{tenbergThermodynamicModelingMRNA2025,walkerAnomalousPhaseBehavior2025,aliLowerCriticalSolution2019,kantasharkerUpperCriticalSolution2019,ohUpperCriticalSolution2024,liSaltInducedLiquidLiquid2021}. Electrostatics therefore establishes one contribution to the underlying thermodynamic driving force, while molecular chemistry determines how that contribution competes with other interactions to produce the observed phase behavior.

In summary, while implicit-solvent models reduce the solvent to a uniform dielectric background and can therefore suggest LCST behavior for charge-containing solutions, our results demonstrate the opposite trend for charged species in a dipolar solvent. Explicitly accounting for ion--dipole coupling reveals a concentration-dependent reversal in the thermal response of electrostatic interactions: isolated pair attractions strengthen with temperature, whereas collective ion--dipole correlations at finite ion concentration cause these interactions to weaken. This reversal produces UCST phase behavior and cannot be inferred from the macroscopic dielectric constant alone, because the local solvent environment governing ion correlations differs substantially from the bulk dielectric response.

More generally, these results show that isolated pair potentials of mean force, although useful for characterizing microscopic interactions, can be qualitatively misleading when used as proxies for bulk phase-separation propensity. Stronger pair attractions upon heating do not necessarily imply LCST behavior once the surrounding ionic environment modifies the interaction itself. Phase behavior in charged fluids is therefore governed not simply by pairwise electrostatic interactions, but by their coupling to collective, solvent-mediated correlations. These findings identify explicit ion--dipole coupling as a key ingredient in temperature-dependent phase behavior and highlight the need to account for solvent degrees of freedom when orientational polarization contributes substantially to the dielectric response\cite{nakamuraIonSolvationLiquid2012,zhuangDissolvesFirstprinciplesTheory2021,beckinghausenInterplayPolymerStructure2023,alonsoExplicitSolventModel2026,buyukdagliExplicitSolventTheory2022}.

This research is supported by funding from Hong Kong Quantum AI Lab, AIR@InnoHK of the Hong Kong Government. We thank Dr. Samuel Varner, Dr. David Beyer and Prof. Alejandro Gallegos for helpful discussions.

\bibliography{references}
\bibliographystyle{apsrev4-2}
\end{document}


\preprint{APS}

\title{Supplemental Material: Explicit Solvents Reverse Temperature Dependence of Phase Behavior in Charged Fluids}

\author{Pierre J. Walker}
 \email{pjwalker@caltech.edu}
\affiliation{Division of Chemistry and Chemical Engineering, California Institute of Technology, Pasadena, CA 91125, United States}

\author{Zhen-Gang Wang}%
 \email{zgw@caltech.edu}
\affiliation{Division of Chemistry and Chemical Engineering, California Institute of Technology, Pasadena, CA 91125, United States
}%



\maketitle

\section{Details on the Theory}
\subsection{Mean-spherical approximation for ions in a dipolar solvent}
\noindent
To describe the phase behavior of monomeric ions in an explicit dipolar solvent, we use the mean-spherical approximation for ion--dipole mixtures (MSA-ID) developed by Blum and Wei\cite{blumAnalyticalSolutionMean1987,weiMeanSphericalApproximation1987}, together with the later reformulation of Simonin and co-workers\cite{simoninSolutionMeansphericalApproximation2020,simoninFullSolutionMean2021}. Here, ion--ion, ion--dipole, and dipole--dipole correlations are treated explicitly, rather than being incorporated through a prescribed dielectric constant.

All species are taken to have the same hard-sphere diameter $\sigma$. The interaction between two ions is
\begin{equation}
u_{ij}^{\mathrm{ion}}(r)
=
\begin{cases}
\infty, & r<\sigma,
\displaystyle
\\\frac{z_i z_j e_c^2}{4\pi\epsilon_0 r},
& r\geq\sigma,
\end{cases}
\end{equation}
where $z_i$ is the valence of ionic species $i$ and $e_c$ is the elementary charge. In contrast to an implicit-solvent description, no dielectric constant appears in the bare Coulomb interaction.

The interaction between an ion and a solvent dipole is
\begin{equation}
u_{ij}^{\mathrm{ion-dipole}}(r)
=
\begin{cases}
\infty, & r<\sigma,\\
\displaystyle
\frac{z_i e_c}{4\pi\epsilon_0 r^3}
\boldsymbol{\mu}_j\cdot\mathbf{r},
& r\geq\sigma,
\end{cases}
\end{equation}
where $\boldsymbol{\mu}_j$ is the solvent dipole vector and $\mathbf{r}$ is the separation vector. Finally, two solvent dipoles interact according to
\begin{equation}
u_{ij}^{\mathrm{dipole-dipole}}(r)=
\begin{cases}
\infty, & r<\sigma,\\
\displaystyle
\frac{1}{4\pi\epsilon_0r^3}
\left[
\boldsymbol{\mu}_i\cdot\boldsymbol{\mu}_j
-
\frac{3(\boldsymbol{\mu}_i\cdot\mathbf{r})
(\boldsymbol{\mu}_j\cdot\mathbf{r})}{r^2}
\right],
& r\geq\sigma.
\end{cases}
\end{equation}

Because dielectric screening arises directly from the dipolar degrees of freedom, it is convenient to introduce the bare Debye screening parameter
\begin{equation}
\kappa_0^2
=
\frac{e_c^2}{\epsilon_0 k_\mathrm{B}T}
\sum_i \rho_i z_i^2,
\qquad
\overline{\kappa}_0=\kappa_0\sigma,
\end{equation}
where the sum extends over the ionic species. The strength of the dipolar coupling is characterized by
\begin{equation}
\chi
=
\frac{\rho_\mathrm{solv}\mu^2}
{3\epsilon_0k_\mathrm{B}T},
\end{equation}
where $\rho_\mathrm{solv}$ and $\mu$ are the number density and dipole moment of the solvent, respectively.

Using an invariant expansion in spherical harmonics\cite{blumInvariantExpansionTwoBody1972,blumInvariantExpansionIII1973}, the MSA-ID solution can be reduced, for equal-diameter species, to three coupled equations for the screening parameter $\Gamma$ and the ion--dipole and dipole--dipole scaling parameters $D$ and $\lambda$:
\begin{widetext}
\begin{align}
&(1+\Gamma\sigma)^2\Gamma^2
+(1-D)(1+\Gamma\sigma)^2
(1+\lambda+\Gamma\sigma)^2
=
\frac{D^2\kappa_0^2}{4},
\label{eq:si_msa_ion}
\\
&\sqrt{D-1},
\frac{1+\Gamma\sigma}{3}
\Big[
D(1+\lambda+\Gamma\sigma)^2
(3+\lambda+\Gamma\sigma)
-(1+\Gamma\sigma)(1+\lambda)
(3+2\lambda+2\Gamma\sigma)
\Big]
=
\frac{D^2\kappa_0\sigma\sqrt{\chi}}{2},
\\
&\frac{9(1+\lambda)^4}{16(2+\lambda)^2}
\left[
(1-\rho_\mathrm{solv}K^{11})^2
+
\frac{(1+\Gamma\sigma)^2(D-1)}{9D^2}
(3+\lambda+\Gamma\sigma)^2
\right]
=
\epsilon_\mathrm{r}.
\label{eq:si_msa_dipole}
\end{align}
\end{widetext}
The auxiliary quantity $K^{11}$ is given by
\begin{align}
1-\rho_\mathrm{solv}K^{11}
&=
\frac{\lambda(2+\lambda)}{3D}
+
\frac{D-1}{3D}
(1+\lambda+\Gamma\sigma)
(3+\lambda+\Gamma\sigma),
\end{align}
and the dielectric constant is related to the dipolar coupling parameter and $\lambda$ through
\begin{equation}
\epsilon_\mathrm{r}
=
1+
\frac{9(1+\lambda)^4}
{16(2+\lambda)^2}\chi.
\label{eq:si_dielectric_msa}
\end{equation}
Thus, $\epsilon_\mathrm{r}$ is not prescribed externally, but follows self-consistently from the solvent density and dipole moment together with the ion-induced changes in the MSA-ID scaling parameters. In particular, changes in ionic concentration alter the dielectric response through their coupling to the solvent orientational degrees of freedom.

The electrostatic contribution to the reduced free-energy density is
\begin{align}
f_\mathrm{ele}
&=
\frac{1}{12\pi}
\Bigg[
\frac{\kappa_0^2}{\sigma}
\left(
\frac{D-1-\Gamma\sigma}{1+\Gamma\sigma}
+
\frac{D-1}{2+\lambda}
\right)
-
\sqrt{\chi(D-1)}
\frac{\kappa_0}{\sigma^2}
\Bigg]
-J',
\label{eq:si_msa_free}
\end{align}
where the full expression for $J'$ involves three contributions:
\begin{equation}
    J' = J'_{\text{ion-ion}}+2J'_{\text{ion-solv.}}+J'_{\text{solv.-solv.}}\,,
\end{equation}
where:
\begin{equation}
    J'_{\text{ion-ion}}=\frac{\pi\sigma^3}{3}\sum_{ij} \rho_i\rho_j(h_{\text{ion-ion},ij}(\sigma))^2\,,
\end{equation}
where:
\begin{equation}
    h_{\text{ion-ion},ij}(\sigma)=\frac{2+\eta}{2(1-\eta)^2}-Z_iZ_j \frac{4l_\mathrm{B,0}}{ D\kappa_0^2\sigma^3}\left(\frac{3(D-1)}{2+\lambda}+(\Gamma\sigma)^2\right)\,.
\end{equation}
For $J'_{\text{ion-solv.}}$:
\begin{equation}
    J'_{\text{ion-solv.}} = \frac{\pi\sigma^3}{3}\sum_{i\in\text{ion}}\sum_{j\in\text{solv.}}\rho_i\rho_j(h_{\text{ion-solv.},ij}(\sigma))^2\,,
\end{equation}
where:
\begin{equation}
    h_{\text{ion-solv.},ij}(\sigma) =-4\sqrt{3} Z_i\frac{\sqrt{\rho_j(D-1)}}{D\sqrt{I}}(\lambda+\Gamma\sigma (1+\lambda+\Gamma\sigma))\,.
\end{equation}
And for $J'_{\text{solv.-solv.}}$:
\begin{equation}
    J'_{\text{solv.-solv.}} = \frac{\pi\sigma^3}{3}\left((h_{\text{solv.-solv.},1}(\sigma))^2+\frac{(h_{\text{solv.-solv.},2}(\sigma))^2}{5}\right)\sum_{ij\in\text{solv.}}\rho_i\rho_j\,,
\end{equation}
where:
\begin{equation}
    h_{\text{solv.-solv.},1}(\sigma) = \frac{Q'^{11}+2q'}{2\sqrt{3}\pi\sigma}\,,
\end{equation}
\begin{equation}
    h_{\text{solv.-solv.},2}(\sigma) = \sqrt{10}\frac{Q'^{11}-2q'}{2\sqrt{3}\pi\sigma}\,,
\end{equation}
and:
\begin{equation}
    Q'^{11}=\frac{2\lambda}{D\rho_\mathrm{solv.}\sigma^3}(\lambda+(D-1)(1+\lambda+\Gamma\sigma))+2\frac{(D-1)}{D}(1+\Gamma\sigma)(1+\lambda+\Gamma\sigma)-\frac{2}{\rho_\mathrm{solv.}\sigma^3}
\end{equation}
\begin{equation}
    q'=-6\frac{(\lambda-1)(\lambda+3)}{(2+\lambda)(1+\lambda)^2}
\end{equation}

The equations recover the appropriate limiting cases. In the absence of solvent dipoles,
\begin{equation}
\chi\rightarrow0,\qquad
D\rightarrow1,\qquad
\lambda\rightarrow1,
\end{equation}
and Eqs.~\eqref{eq:si_msa_ion}  reduce to the primitive-model MSA with a vacuum dielectric background, $\epsilon_\mathrm{r}=1$. In this limit,
\begin{equation}
J'=\frac{\Gamma^3}{3\pi}.
\end{equation}
Conversely, in the ion-free limit,
\begin{equation}
\kappa_0=2\Gamma=0,
\qquad
D=1,
\end{equation}
and Eq.~\eqref{eq:si_msa_dipole} reduces to
\begin{equation}
\frac{\lambda^2(1+\lambda)^4}{16}
=
\epsilon_\mathrm{r}.
\end{equation}
Together with Eq.~\eqref{eq:si_dielectric_msa}, this recovers Wertheim's MSA result for the dielectric constant of a dipolar hard-sphere fluid\cite{wertheimExactSolutionMean1971}. For each temperature and composition, Eqs.~\eqref{eq:si_msa_ion}--\eqref{eq:si_msa_dipole} are solved simultaneously for $\Gamma$, $D$, and $\lambda$. Numerically, we initialize the solution using the corresponding ion-free and dipole-free limits, with $D$ initially set to unity.

\subsection{Phase equilibrium}
\noindent
To match the simulations, the total number density is fixed at
\begin{equation}
\rho\sigma^3=0.8.
\end{equation}
We consider a symmetric monovalent electrolyte, for which local electroneutrality requires
\begin{equation}
\rho_+=\rho_-=\frac{\rho_\mathrm{ion}}{2}.
\end{equation}
Because all particles have the same diameter, the charged-particle volume fraction may be written as
\begin{equation}
\phi=\frac{\rho_\mathrm{ion}}{\rho},
\end{equation}
with solvent fraction
\begin{equation}
\phi_\mathrm{solv}=1-\phi.
\end{equation}
At fixed total density, the system has a single independent composition variable $\phi$. Two phases with compositions $\phi^\alpha$ and $\phi^\beta$ coexist when their chemical potentials and pressures are equal, equivalently satisfying the common-tangent conditions
\begin{align}
\left.
\frac{\partial f}{\partial\phi}
\right|_{\phi^\alpha,T}
&=
\left.
\frac{\partial f}{\partial\phi}
\right|_{\phi^\beta,T},
\\
f(\phi^\alpha)
-
\phi^\alpha
\left.
\frac{\partial f}{\partial\phi}
\right|_{\phi^\alpha,T}
&=
f(\phi^\beta)
-
\phi^\beta
\left.
\frac{\partial f}{\partial\phi}
\right|_{\phi^\beta,T}.
\end{align}
The critical point satisfies
\begin{align}
\left(
\frac{\partial^2f}{\partial\phi^2}
\right)_T
&=0,
\\
\left(
\frac{\partial^3f}{\partial\phi^3}
\right)_T
&=0.
\end{align}

All MSA-ID and phase-equilibrium calculations are performed using Clapeyron.jl\cite{walkerClapeyronjlExtensibleOpenSource2022}, with composition derivatives evaluated using ForwardDiff.jl\cite{revelsForwardModeAutomaticDifferentiation2016}.

\section{Additional Results}

\subsection{Bjerrum length at co-existence}
\begin{figure}[ht!]
    \includegraphics[width=0.5\linewidth]{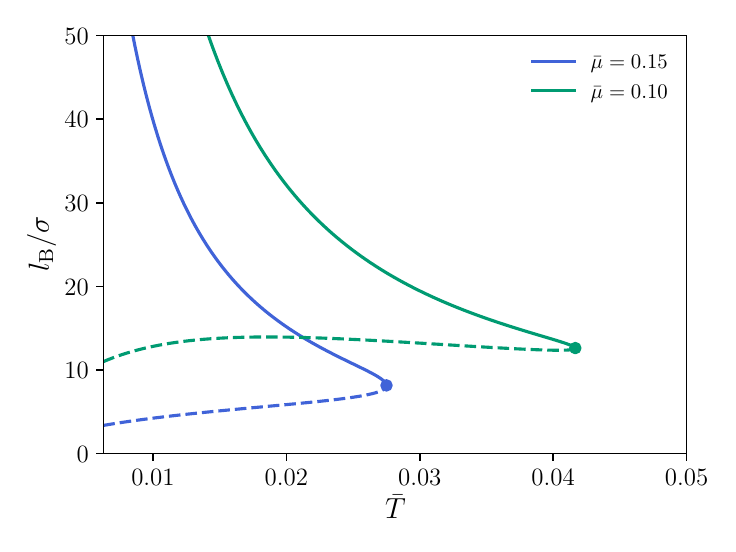}
    \vspace{-0.5cm}
    \caption{Bjerrum length $l_\mathrm{B}/\sigma$ in the charge-concentrated (solid curves) and charge-dilute (dashed curves) as a function of dimensionless temperature $\bar{T}$ along the co-existence phase boundary at two dipole strengths $\bar{\mu}$. All results obtained from MSA-ID.}
    \label{fig:3}
\end{figure}
In Fig.~\ref{fig:3}, we examine the temperature dependence of the Bjerrum length in the charge-dilute and charge-concentrated phases along the coexistence boundary for two solvent dipole strengths. The two phases exhibit markedly different trends. In the charge-dilute phase, $l_\mathrm{B}$ is smaller and varies only weakly with temperature, increasing slightly for the stronger dipole strength, $\bar{\mu}=0.15$. In contrast, the charge-concentrated phase consistently exhibits a decaying Bjerrum length with increasing temperature. Comparing these trends with the phase diagrams in the main text shows that the thermal response of $l_\mathrm{B}$ in the charge-concentrated phase correlates more closely with the observed phase behavior than does that of the charge-dilute phase.

\bibliography{references}